\documentstyle[preprint,aps,epsf]{revtex}
\begin{document}
\tightenlines
\draft
\title{\bf Self-Organisation to Criticality in a 
System without Conservation Law}

\author{Stefano Lise}

\address{Department of Mathematics, Huxley Building, Imperial College
of Science, Technology, and Medicine, London UK SW7 2BZ \\}
\date{\today}

\maketitle 

\begin{abstract}
We numerically investigate the approach to the stationary state in the
nonconservative  Olami-Feder-Christensen (OFC) model for earthquakes. 
Starting from  initially random configurations, we monitor the average 
earthquake size in different portions of the system as a function of time 
(the time is defined as the input energy {\em per site} in the system). 
We find that the process of 
self-organisation develops from the boundaries of the system and it is 
controlled by a dynamical critical exponent $z\simeq 1.3$ that appears 
to be universal over a  range of dissipation levels of the local dynamics. 
We show moreover that the transient time of the system $t_{tr}$ scales with 
system size $L$ as $t_{tr} \sim L^z$. We argue that the (non-trivial) scaling 
of the transient time in the OFC model is associated to the establishment of 
long-range spatial correlations in the steady state.  
\end{abstract}

\vspace{0.3cm}


\section{Introduction}

The idea of self-organised criticality (SOC) was introduced by Bak, Tang and
Wiesenfeld (BTW)~\cite{btw} as a possible paradigm for the widespread 
occurrence in nature of scale free phenomena. It refers  to the intrinsic 
tendency of extended, non-equilibrium systems to spontaneously self-organise 
into a dynamical critical state. In general, SOC systems are driven externally
at a very slow rate and relax with bursts of activity (avalanches) on a very 
fast (almost instantaneous) time scale.
The standard signature of SOC is a power law distribution of avalanche sizes 
and in this sense is the system said to be critical. Typical physical 
realizations of this phenomena includes, among others, earthquakes, 
forest fires and biological evolution (for a review, see e.g., 
Ref.~\cite{bak_book,jen_book}). 

A number of simple lattice models have been developed to test the 
applicability of SOC to a variety of complex interacting dynamical 
systems~\cite{bak_book,jen_book}. 
In general these models reach a stationary critical state after a 
sufficiently long transient time. Not much attention though has been paid 
to the self-organisation process and studies have mainly concentrated on the 
properties of different models at stationarity. 
This is partly justified by the fact that some of the most studied models, 
such as the BTW~\cite{btw} and the Zhang~\cite{zhang} models, display a 
very simple transient time behaviour. Indeed in these models the relaxation 
time does not scale with system size (time is defined as the input energy per 
site in the system)~\cite{achille}. On the other hand, there are also models 
with a more complex behaviour (see,  e.g., Ref.~\cite{sibani}).

In this paper we investigate the approach to the stationary state in 
the so-called Olami-Feder-Christensen (OFC) model for earthquake 
dynamics~\cite{ofc}. 
This model has in recent years attracted a considerable deal of attentions 
especially because it has been proposed as an example of a system displaying 
SOC behaviour even with a non-conservative 
dynamics~\cite{grassberger,socolar,middleton,ceva}. 
One of the most important question in this field is indeed whether a conserving
local dynamics is a necessary condition for SOC~\cite{hwa,grinstein}. For 
example, it is well known that the BTW model is subcritical if dissipation is 
introduced~\cite{manna}. The presence of criticality in the non-conservative 
OFC model is still debated~\cite{carvalho,kim}. Recent numerical 
investigations, though, have shown that the model on a square lattice displays
scaling behaviour, up to lattice sizes presently accessible by computer 
simulations~\cite{lisepac1,lisepac2}.

The present investigation complements previous analysis of the OFC model which
were based on the study of the probability distribution for earthquake sizes. 
It provides further numerical support in favour of criticality in the 
non-conservative regime. Indeed, we will show that the model displays a non 
trivial transient time behaviour: the relaxation time scales with system 
size and it is controlled by a dynamical critical exponent $z$ that appears 
to be universal  over a  range of dissipation levels of the local dynamics.
Moreover we will establish the presence of long range spatial correlations in 
the system. In so doing, we will be able to gain some insight into the 
mechanisms behind criticality in non-conserving systems, mechanisms that are 
very different from those at work in systems with a conservation law.

The plan of the paper is as follows. In section II we describe the model
and briefly summarise previous foundings relevant to our investigation. 
In section III we define the quantities of interest and present the results
of our numerical study. Finally, in section IV we discuss our main 
conclusions.

\section{The model}

The OFC model is a coupled map lattice model, where to each site $(i,j)$ 
of a square lattice of linear size $L$ is associated a real variable 
$F_{ij}$. In the initial state, at time $t=0$, the values of the $F_{ij}$
are chosen randomly in the uniform interval $(0,F_{c})$.
Subsequently the variables evolve according to  the following two-steps 
dynamical rules:  (i)
if all sites in the system are stable (i.e., $F_{ij} < F_{c}$), they 
increase simultaneously and uniformly at a constant rate
\begin{equation}
\label{slow_dyn}
\frac{\partial F_{ij}(t)}{\partial t}=v  ;
\end{equation}
(ii) as soon as one of them reaches the threshold value $F_{c}$, the 
uniform driving is stopped and an ``earthquake'' starts:
\begin{equation}
\label{fast_dyn}
F_{ij} \ge F_{c} \Rightarrow \left\{ \begin{array}{l}
                                       F_{ij} \rightarrow 0 \\
                         F_{nn} \rightarrow F_{nn} + \alpha F_{ij}
                                      \end{array} \right.
\end{equation}
where ``$nn$'' denotes the set of nearest neighbour sites of $(i,j)$ and 
$\alpha$ is a parameter that controls the level of conservation of the 
dynamics ($\alpha=1/4$ corresponds to the conservative case).
The ``toppling'' rule  (\ref{fast_dyn}) can possibly produce a chain reaction,
which ends when there are no more unstable sites in the system. At that point,
the uniform growth (\ref{slow_dyn}) starts again. 
In the following we will assume,  without loss of generality, a unit growth 
rate, i.e. $v=1$. 
A crucial point in the description of the model is the choice of boundary 
conditions and, in accordance with previous investigations, we will consider 
open boundary conditions.
These conditions imply that sites close to the boundaries topple according to 
(\ref{fast_dyn}) but have a smaller coordination number.

There is a clear separation of time scales in the system: earthquakes occur 
instantaneously on the slow time scale of the driving. 
The time in the system is therefore set by the slow time variable $t$. 
By construction, moreover, the time coincides with the input energy {\em per
site} in the system. We will make use of this latter observation  when we
will try to compare the behaviour of the OFC model with other models, such
as the BTW model or the Zhang model.

After a sufficiently long transient time, the system settles into a 
stationary state, where the statistical properties of the model (e.g. the
probability distribution for earthquake sizes) do not depend on time.  
In the BTW model and in the Zhang model, the transient time is relatively 
brief and does not scale with system size. On average, an input energy 
proportional to system size is needed to reach the steady 
state~\cite{achille}. 
On the contrary, transient times in the OFC model are known to be extremely 
long, especially for large lattices.
For example, it was claimed in Ref.~\cite{grassberger} that $4 \times 10^8$ 
earthquakes are not enough to reach stationarity in a system of size $L=200$ 
for $\alpha=0.1$. 
This conclusion was reached by observing the very slow convergence of the 
mean earthquake size to an asymptotic value during the transient time. 
In Ref.~\cite{grassberger} though, a  systematic investigation of the 
relaxation times for different $\alpha$ and $L$ was not attempted.
A more quantitative approach to the problem was proposed in 
Ref.~\cite{middleton} by Middleton and Tang (MT).
According to these authors a ``self-organised'' region develops first close 
to the boundaries and propagates thereafter into the bulk of the system. 
The distance from the boundaries of the invasion front grows with time 
as a power law, $d(t) \sim t^{\gamma (\alpha)}$, with 
$\gamma = 0.23 \pm 0.08, 0.63 \pm 0.08$ for $\alpha =0.07, 0.15$,  
respectively. The system reaches stationarity when the SOC region crosses
the whole sample ($d(t) \sim L$). Assuming that the power law behaviour
of $d(t)$ holds till saturation, than the transient time
of the system should scale as $t_{tr} \sim L^{1/\gamma (\alpha)}$.
More recently, it has been suggested in Ref.~\cite{ceva} that two distinct 
relaxation times exist in the system, associated respectively with the power 
law region and the ``tail'' (induced by finite size effects)  of the 
distribution of earthquake sizes. According to this study, the former should 
stabilise much faster than the latter.

\section{Results}

\begin{figure}[hb]
\narrowtext
\epsfxsize=2.5in
\centerline{\epsffile{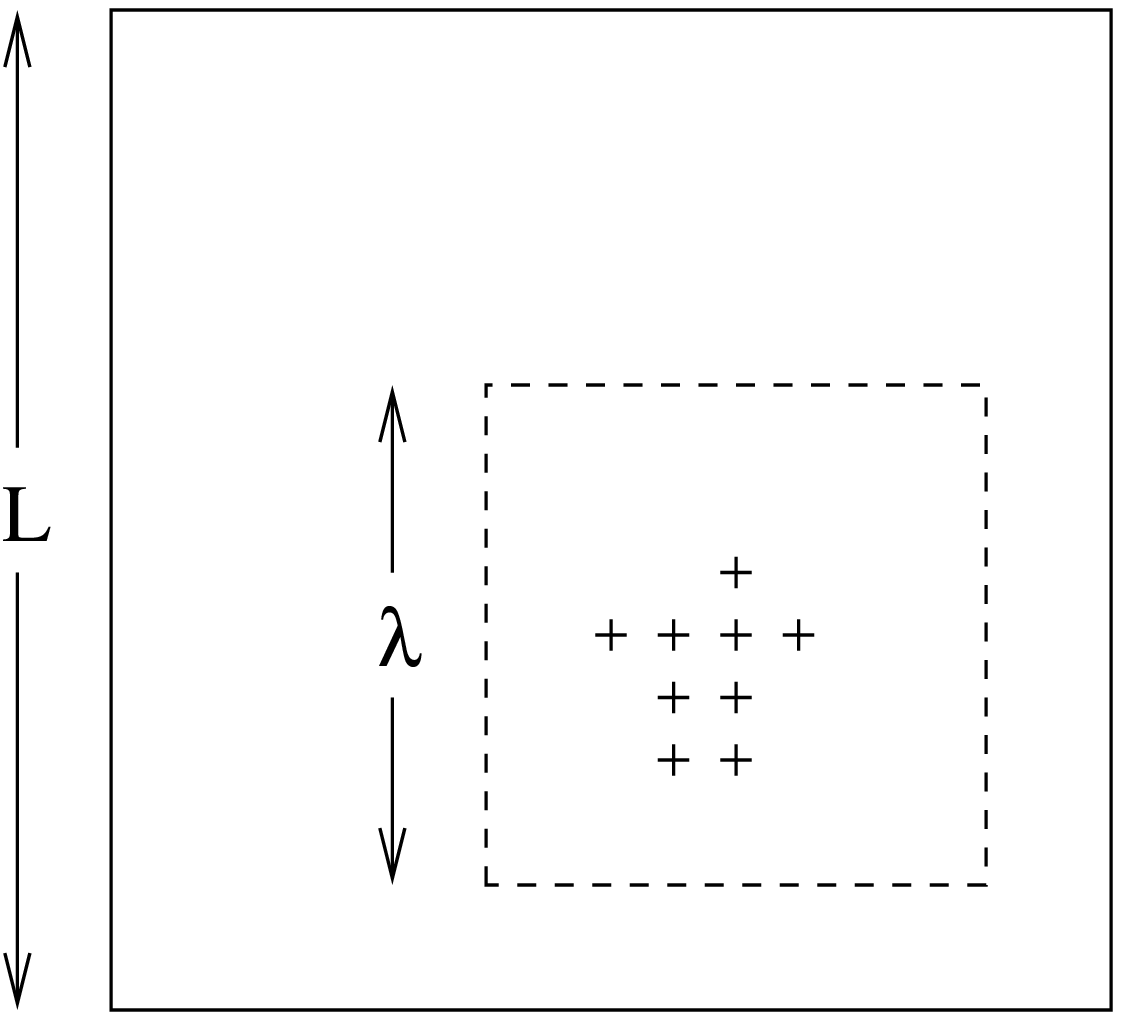}}
\protect\vspace{0.2cm}
\caption[1]{\label{scheme}
Schematic representation of earthquakes entirely confined within a subsystem 
of linear size $\lambda$ (dashed line). Toppling sites are denoted with a 
cross.
}
\end{figure}

Most of the studies on the OFC model at stationarity have concentrate 
on the probability distribution of earthquake sizes, $P_L (s)$, where $L$ is
the size of the system and $s$ is the total number of sites that topple 
during an earthquake. This probability distribution does not show
simple finite size scaling, at least in the range of lattice sizes accessible 
to simulations at present~\cite{lisepac1}.  
In a recent paper~\cite{lisepac2}, we have focused instead on the properties
of earthquakes confined within a fictitious subsystem of linear size 
$\lambda$ (see fig.~\ref{scheme}). The model is driven according to its 
usual dynamics but only those particular earthquakes that are entirely 
contained within the subsystem are counted. We have shown that if
$\lambda$ is sufficiently smaller than $L$ the size distribution for this 
subset of earthquakes, $P_{conf} (\lambda,s)$, obeys ordinary finite size 
scaling, i.e. $P_{conf} (\lambda,s)\simeq \lambda^{-\beta}f(s/\lambda^D)$, 
where the exponents $\beta=3.6$ and $D=2$ are universal over a range of 
values of $\alpha$. 

In this work we want to address the issues briefly summarised in the 
previous section concerning the approach to stationarity in the OFC model. 
In order to be able to formulate scaling hypothesis and make use of 
collapse plots, we will proceed in a way similar to that of 
Ref.~\cite{lisepac2}. We will consider earthquakes localised within given 
subsystems, in particular subsystems placed (a) at the boundaries and (b) at 
the centre of the system. 
As it is a prohibitive task to determine the time evolution of the entire 
distribution $P_{conf} (\lambda,s)$, we will restrict ourselves to the mean 
earthquake size, $<s>_{\lambda,L} (t)$, and, in general, to  $q$-th moment 
(up to $q=4$) of the distribution, $<s^q>_{\lambda,L} (t)$.  
To determine numerically these quantities we have run several simulations 
with different initial conditions, partitioning the time into bins of size 
$\Delta t$. Let $n$ be the number of earthquakes occurring
between time $t-\Delta t/2$ and  $t+\Delta t/2$ in a given realization of
the system and let $s_1, \ldots, s_n$ be the sizes of these earthquakes.
Then we define
\begin{equation}
<s^q>_{\lambda,L} (t) \simeq
\frac{<\left[ s_1^q + \ldots +  s_n^q \right]_{t-\Delta t/2}^{t+\Delta t/2}>}
     {<\left[ n \right]_{t-\Delta t/2}^{t+\Delta t/2}>} 
\end{equation}
where $< \ldots>$ denotes an average over different realizations of the 
system, i.e. over different initial conditions. For each system, the parameter 
$\Delta t$ has been choose small with respect to the transient time 
$t_{tr}$ but large enough to collect reasonably accurate statistics.

We consider first the case of subsystems placed adjacent to a boundary of 
the system, in a symmetric position with respect to the corners.
In figure 2 we report $<s>_{\lambda,L}$ as a function of time for 
$\alpha =0.18$ and some $\lambda$ and $L$. We observe that if the linear 
dimension $\lambda$ of the box is sufficiently smaller than the linear 
dimension of the system $L$ (approximately $L \ge 4 \lambda$) then the curve 
$<s>_{\lambda,L} (t)$ becomes indistinguishable for different $L$.
This has been verified also for other values of $\alpha$ and $\lambda$.
We will therefore denote with  $<s>_{\lambda} (t)$ the mean earthquake size 
in this limit.
It is already visible from figure 2 that the relaxation time of
$<s>_{\lambda} (t)$ increases with $\lambda$. This is an indication in support
of the scenario proposed by MT. Regions close to the boundaries reach 
stationarity sooner, signalling that an invasion front is moving toward the
bulk of the system. Some of MT conclusions nonetheless will have to be 
modified as we will show later.

In order to describe quantitatively the invasion from the boundaries of
the self-organised region we make the following simple scaling hypothesis
\begin{equation}
\label{bound_scal}
<s>_{\lambda} (t)= \lambda^{\eta} F(t/\lambda ^z)
\end{equation}
where $\eta$ and $z$ are two suitable critical exponents. In particular
$z$ is a dynamical critical exponent that should satisfy $z=1/\gamma$, where
$\gamma$ is the ``invasion'' exponent as defined by MT. 
In the limit of $t \rightarrow \infty$ the scaling function $F(x)$ saturates 
to a constant, implying that the exponent $\eta$ is related to the finite
size exponents of the probability distribution $P_{conf}(\lambda,s)$ by the 
relation  $\eta =2 D -\beta \approx 0.4$. In figure \ref{fig_3} we report 
collapse plots of the form (\ref{bound_scal}) for various values of $\alpha$.
We observe that a reasonably good collapse could be obtained for all the
$\alpha$ if we choose the universal exponent $z=1.3 \pm 0.1$. We are therefore
led to conclude that a universal exponent $z$ exist contrary to the claims 
by MT. 

We consider next subsystems of different sizes placed at the centre of the 
system. We observe that the relaxation time does not depend on the size of 
the subsystems (see fig.~\ref{fig_4}a).
This confirm that the self-organisation mechanism develops from the 
boundaries and that the system enters stationarity when the self-organised
region span the whole system. Only when the bulk of the system is reached by 
the self-organised region is stationarity settled so that concentric
subsystem of different sizes will inevitably reach stationarity at the same 
time.
We have also tested the relaxation times for higher moments of the avalanche
probability distribution to see whether different parts of the distribution
(e.g. power law part and the ``tail'') have different relaxation times as
suggested in Ref.~\cite{ceva}. In our investigation though we have not 
observed any significant difference in the relaxation times associated with 
different moments. We report as an example in fig.~\ref{fig_4}b the 
behaviour of the first, second and fourth moment in a particular case.

The scaling equation (\ref{bound_scal}) suggests that the transient time in 
the OFC model scales with system size as $t_{tr} \sim L^z$. One way to test 
this is by comparing the time behaviour of the average earthquake size in a 
central subsystem of size $\lambda$ for different system sizes $L$. Indeed the
asymptotic value $<s>_{\lambda,L} (t \rightarrow \infty)$ should not depend on
$L$. We report as example the case for $\alpha=0.18$ in fig.~\ref{fig_5}, 
where we have rescaled the time by a factor $L^z$. The curves show some noisy 
behaviour, due to the difficulties in 
collecting good statistics (relatively few earthquakes occur in the bulk 
of the system as compared to the boundaries). Nonetheless the value deduced 
for the exponent, $z \approx 1.3$, is consistent with the determination made 
through the analysis of the earthquakes occurring at the boundaries. We have
obtained similar results also for other values of $\alpha$. In addition, 
besides the average earthquake size, we have considered also the time 
behaviour of other quantities such as the roughness of the energy landscape 
(in analogy to surface growth problems) and the number of earthquakes per 
unit time. All these different quantities on average reach stationary at the 
same time.

The algebraic divergence of the relaxation time with system size reflects 
the presence of long-range spatial correlations in the stationary state. 
Indeed if correlations were only short range, than one would expect that the 
transient time would not scale with system size. This is for example the case 
for the BTW model in $d$ dimensions, where the height-height correlations are
algebraic but decays as fast as $r^{-2d}$ ($r$ being the distance between two
sites)~\cite{majumdar}. We have measured for various $L$ and $\alpha$ the 
probability distribution of the spatially averaged force in the system
\begin{equation}
\label{M}
M=\frac{1}{L^2} \sum_{i,j=1}^{L} F_{i,j} 
\end{equation}
In a system with sufficiently short range correlations, this probability 
distribution would tend, in the limit $L \rightarrow \infty$, to a gaussian 
distribution around the mean due to the central limit theorem (this is indeed 
what results in the BTW model). Let $<M>$ and $\sigma _M$ be respectively 
the average and the standard deviation of the distribution. In figure 
\ref{fig_6} we have plotted $\log(\sigma _M P(M))$  versus 
$(M-<M>)/\sigma _M$ for various $L$ and $\alpha$. Using these 
coordinates a gaussian function would result in an inverse parabola. 
For each $\alpha$ value the data of figure \ref{fig_6} collapse on a single 
function, which is clearly not gaussian (deviations from gaussianity are more
pronounced for increasing $\alpha$ values). This indicates that the central 
limit theorem does not hold in this case, not even for large $L$, suggesting 
that long range algebraic correlations are present and therefore the 
sum (\ref{M}) can not be decomposed into a sum of independent terms. 
This observation is in agreement with the results reported in 
Ref.~\cite{grinstein}  where the presence of long-range spatial correlations 
were deduced from the behaviour of a suitably defined susceptibility, 
$\chi \equiv (L \sigma _M)^2$. It was claimed that $\chi$ diverges as $L^2$ 
and correspondingly that $\sigma _M$ is, to leading order, independent on $L$ 
(if $M$ was a sum of uncorrelated variables, $\sigma _M$ would decrease as 
$1/L$).
In our investigation we have found that  $\sigma _M$ slightly increases with 
$L$ but asymptotically tends to a constant value,  in accordance with 
Ref.~\cite{grinstein} ($M$ is a bounded variable so $\sigma _M$ can not grow 
indefinitely).

\section{Conclusions}

In conclusion, in this paper we have examined the process of self-organisation
in the OFC model. By considering earthquakes confined within a given subsystem
we have been able to clarify some of the issues related to this problem. In 
accordance with Middleton and Tang~\cite{middleton} we have found that SOC 
develops first close to the boundaries and subsequently invades the 
interior of the system. The invasion process is controlled by a dynamical
critical exponent, $z \simeq 1.3$, which, contrary to previous claims, is
universal over a range of values of the dissipation level of the local 
dynamics. 
We have shown moreover that the transient time in the system scales with
system size as $t_{tr} \sim L^z$. This is a peculiarity of the OFC model as
other ``sandpile-like'' models (e.g. the BTW and the Zhang models) do not
display any scaling in the transient time. We have associated this feature
with the presence of long-range spatial correlations in the stationary state. 

Our findings are in general agreement with recent works on the OFC 
model~\cite{lisepac1,lisepac2}. Indeed we have provided complementary evidences
(not based on the probability distribution for earthquake sizes) that the
model is critical even in a non conservative regime. Moreover it confirms
that there is universality in the system and that finite-size scaling can be 
recovered by considering subsystems whose linear extent is sufficiently small.

Finally, it is interesting to remark that the probability distribution for the
spatially averaged force in the system is somewhat reminiscent of a 
probability distribution observed in a confined turbulent flow 
experiment~\cite{bramwell} (BHP). As a term of comparison we have reported
in fig.~\ref{fig_6} the BHP functional form over-imposed to the curve for
$\alpha=0.21$. Attempts to link SOC systems to turbulent phenomena have long 
been suggested, but only recently  this has been put on a firmer 
basis~\cite{mario}.

\medskip
 
We thank M.~Paczuski for helpful conversations and H.J.~Jensen for a critical
reading of the manuscript. This work was supported by the EPSRC (UK),
Grant No. GR/M10823/01 and Grant No. GR/R37357/01.

\begin{figure}[hb]
\narrowtext
\epsfxsize=4.in
\centerline{\epsffile{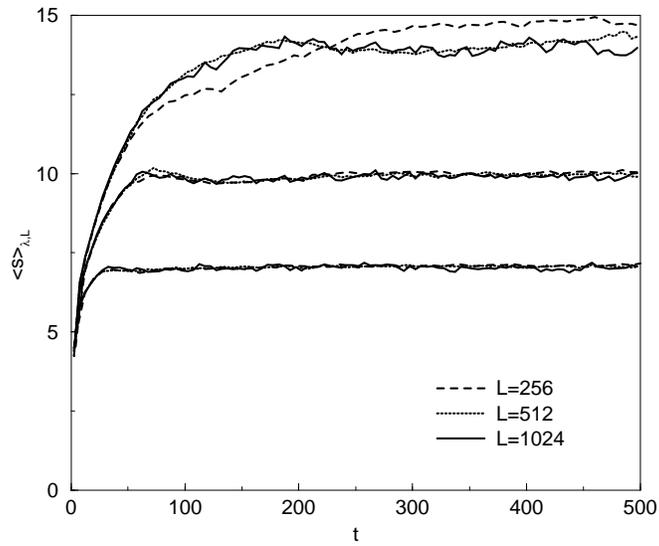}}
\caption[1]{\label{fig_2}
Average earthquake size $<s>_{\lambda,L}$ in a subsystem placed at the 
boundary as a function of time; $\alpha =0.18$ and, from bottom to top, 
$\lambda=32$, $\lambda=64$ and  $\lambda=128$.
}
\end{figure} 
                                                                   
\begin{figure}[hb]
\narrowtext
\epsfxsize=4.5in
\centerline{\epsffile{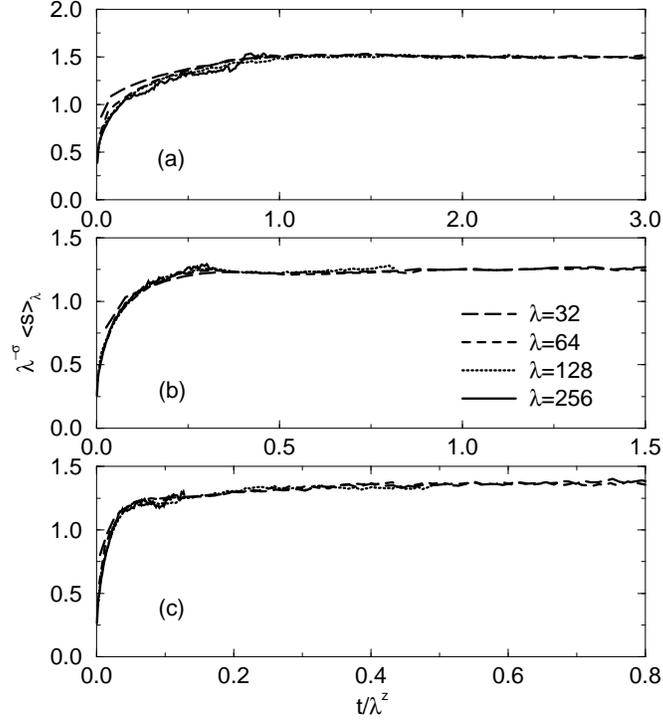}}
\caption[1]{\label{fig_3} Collapse plots of $<s>_{\lambda} (t)$ for a 
subsystem placed at the boundary and for (a) $\alpha=0.15$, (b) $\alpha=0.18$
and (c) $\alpha=0.21$. The value of the dynamical critical exponent is 
$z=1.3$.
}
\end{figure} 

\begin{figure}[hb]
\narrowtext
\epsfxsize=4.in
\centerline{\epsffile{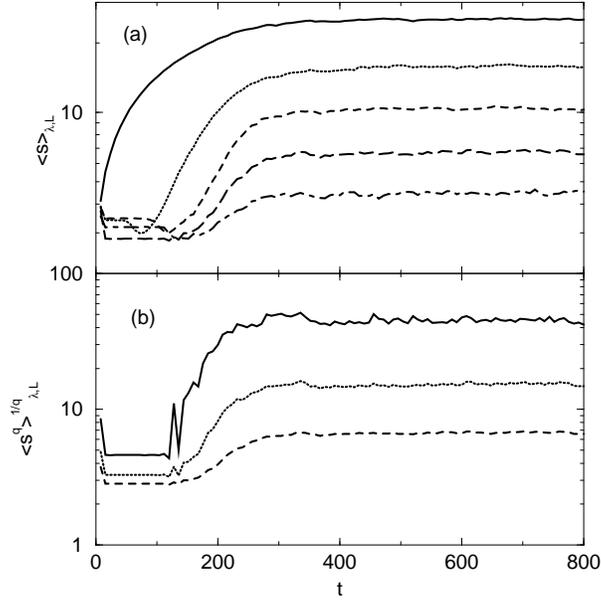}}
\caption[1]{\label{fig_4} Time dependence of the average earthquake sizes in 
subsystems placed at the centre of the system for  $\alpha=0.18$ and $L=256$; 
(a) average earthquake size $<s>_{\lambda,L}$ for, from bottom to top, 
$\lambda=16,32,64,128,256$ and (b) $q-th$ moment of the distribution for 
$\lambda=32$ and, from bottom to top, $q=1,2,4$.

}
\end{figure} 
 
\begin{figure}[hb]
\narrowtext
\epsfxsize=4.in
\centerline{\epsffile{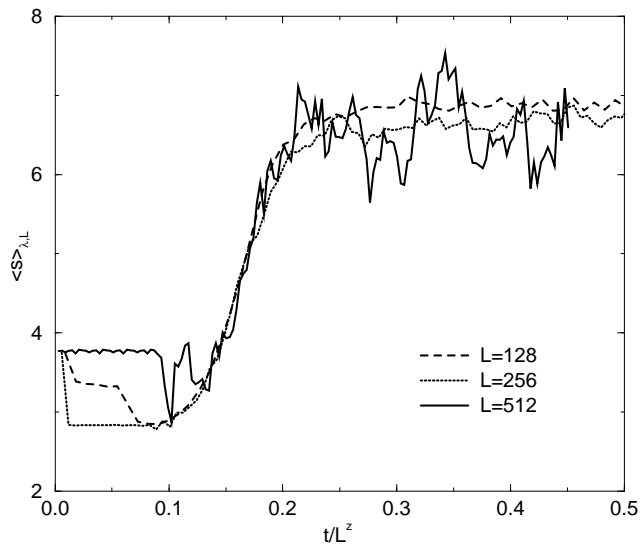}}
\caption[1]{\label{fig_5}
Collapse plots of $<s>_{\lambda, L} (t)$ for a subsystem of size $\lambda=32$
placed at the centre of a system of size $L$; the conservation parameter is 
$\alpha=0.18$. The value of the dynamical critical exponent is $z=1.3$.
}
\end{figure} 

\begin{figure}[hb]
\narrowtext
\epsfxsize=4.in
\centerline{\epsffile{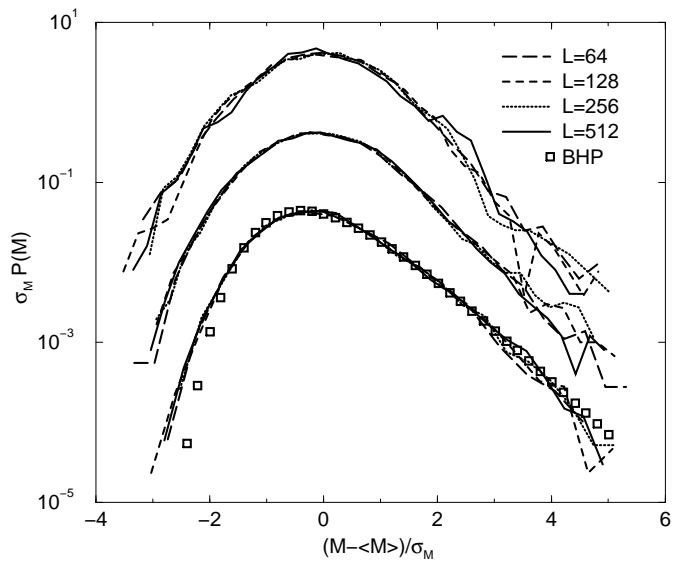}}
\caption[1]{\label{fig_6}
Rescaled probability distribution of the spatially averaged force in the system
for, from bottom to top, $\alpha=0.21, 018, 0.15$. $<M>$ and $\sigma _M$ are
respectively the average and the standard deviation of the distribution. The
top and bottom curves have been shifted by a factor of $10$ respectively up 
and down for visual clarity. Squares represent the BHP (rescaled) probability 
distribution observed in an experiment on turbolence~\protect\cite{bramwell}.}
\end{figure}

\end{document}